\documentclass[twocolumn,apj]{emulateapj}

\newcommand{\bb}{\bm{B}}

\usepackage{color}
\usepackage{bm}
\usepackage{amsmath,amssymb}

\begin{document}

\newcommand{\red}[1]{{\color{red}{\textit{#1}}}}
\newcommand{\blue}[1]{{\color{blue}{\textit{#1}}}}

\title{Nonlinear force-free modeling of flare-related magnetic field changes at the photosphere and chromosphere}

\author{Lucia Kleint\altaffilmark{1,2}}
\altaffiltext{1}{University of Applied Sciences and Arts Northwestern Switzerland, Bahnhofstrasse 6, 5210 Windisch, Switzerland}
\altaffiltext{2}{Kiepenheuer-Institut f\"ur Sonnenphysik, Sch\"oneckstrasse 6, 79104 Freiburg, Germany}
\author{Michael S. Wheatland\altaffilmark{3}}
\altaffiltext{3}{Sydney Institute for Astronomy, School of Physics, University of Sydney, NSW 2006, Australia}
\author{Alpha Mastrano\altaffilmark{3}}
\author{Patrick I. McCauley\altaffilmark{3}}

\begin{abstract}
Rapid and stepwise changes of the magnetic field are often observed during flares but cannot be explained by models yet. Using a 45 min sequence of SDO/HMI 135 s fast-cadence vector magnetograms of the X1 flare on 2014-03-29 we construct, at each timestep, nonlinear force-free models for the coronal magnetic field. Observed flare-related changes in the line-of-sight magnetic field $B_{\rm LOS}$ at the photosphere and chromosphere are compared with changes in the magnetic fields in the models. We find a moderate agreement at the photospheric layer (the basis for the models), but no agreement at chromospheric layers. The observed changes at the photosphere and chromosphere are surprisingly different, and are unlikely to be reproduced by a force-free model. The observed changes are likely to require a change in the magnitude of the field, not just in its direction. 
\end{abstract}

\section{Introduction}

While photospheric magnetic field measurements are readily available, measurements at the solar chromosphere and in the corona are less common and less reliable. Often fields in these higher atmospheric layers are approximated by modeling, particularly by non-linear force-free field (NLFFF) extrapolations. Our goal is to test the agreement of such extrapolations with chromospheric observations, particularly whether the models reproduce the magnetic field changes that are observed in the photosphere and in the chromosphere during a flare.

The equations of the NLFFF model may be written as

\begin{equation}
\nabla\cdot\bb=0,\label{eq:nlfff2}
\end{equation} 
and
\begin{equation}
 \nabla\times\bb=\alpha\bb\label{eq:nlfff1}
\end{equation}
where $\bb$ is the magnetic field vector, and $\alpha$ is the force-free parameter. {{Equation (\ref{eq:nlfff2}) states the fundamental physical condition that the magnetic field must be divergence-free, while Equation (\ref{eq:nlfff1}) states our assumption that the Lorentz force in the corona is zero}}. NLFFF extrapolations use photospheric vector magnetogram data as boundary conditions
to reconstruct the coronal magnetic field \citep[e.g.,][]{2012LRSP....9....5W}. {However, a given set of photospheric observations over-determine the force-free model. The data present two different choices for the boundary conditions, implying two possible solutions to the model. Hence additional choices must be made in the modeling, and the results depend to some extent on the specific choices made \citep[e.g.,][]{2015ApJ...811..107D}.}

A flare is attributed to a change in the coronal magnetic field configuration, involving magnetic reconnection and leading to a release of energy. Free energy stored in the solar coronal field is converted, for example, into particle acceleration and heating of the solar atmosphere. Observations often show abrupt and permanent changes of photospheric magnetic fields during flares
 \citep[e.g.,][]{wang1992,wangetal1994,kosovichevzharkova1999, cameronsammis1999, kosovichevzharkova2001,sudolharvey2005,petriesudol2010}, but their mechanism is not yet fully understood. Photospheric magnetic fields preferentially change near the polarity inversion line, and the line-of-sight magnetic field is equally likely to increase or decrease \citep{castellanosetal2018}. Studies with vector magnetograms show that the horizontal field tends to increase close to the neutral line, in a direction parallel to the neutral line \citep[e.g.][]{2012ApJ...759...50P}. In contrast, chromospheric magnetic field changes are more difficult to study because of the lack of continuous space-based chromospheric polarimetric observations and because of the more complex interpretation of chromospheric spectral lines. \citet{kleint2017} recently reported observations of chromospheric magnetic field changes during the X1 flare on 2014-03-29, which demonstrated a surprising disparity between the photosphere and the chromosphere. Changes in the magnetic field at the chromosphere were observed to occur over larger areas than at the photosphere, the changes were stronger, and in many cases their locations, sign, and timing did not coincide with those at the photosphere.

This leads to the question of whether NLFFF extrapolations are able to reproduce and explain photospheric and chromospheric magnetic field changes during flares. In this paper we address this question, by examining again the data from the X1 flare SOL2014-03-29T17:48.

{We note that there are two basic limitations of NLFFF modeling for our purpose. First, the nonlinear force-free model does not represent accurately the photosphere-chromosphere transition region, because it excludes non-magnetic forces. Second, the static NLFFF model cannot represent the dynamic fields present during the flare. To address the second problem, we construct a long sequence of NLFFF reconstructions starting before the flare and ending after. This sequence is used to identify permanent, flare-related changes in the NLFFF models.}
  
\section{Observations and data reduction}\label{obs}

The X1 flare SOL20140329T17:48, the famous ``best-observed'' flare \citep[e.g.][]{kleintetal2015, judgeetal2014, battagliaetal2015, kleintetal2016, rubiodacostaetal2016} was observed by the ground-based Dunn Solar Telescope (DST) and most current solar spacecraft. Its location near disk center (heliocentric angle $\mu$=0.8) in AR 12017 and the availability of spectroscopy and polarimetry in multiple wavelengths make it a well-suited target for studies.

For the observational comparison to NLFFF extrapolations, we used the data from \citet{kleint2017} who determined the permanent and stepwise magnetic field changes by fitting an arctan function \citep{petriesudol2010} to photospheric and chromospheric line-of-sight magnetic field data ($B_{\rm LOS}$) after removing solar rotation by cross-correlation of each subsequent frame. 

For the photospheric changes, the data series hmi.M\_45s\_nrt from 
the Helioseismic and Magnetic Imager (HMI) onboard the Solar Dynamics Observatory (SDO) \citep{scherrerhmi2012} was used with a fitting time range of 30 minutes (17:30-18:00 UTC) and a cadence of 45 seconds. Additionally, as a cross-check, we carried out the same fitting on $B_{\rm LOS}$ constructed from HMI High-Cadence Vector Magnetograms, which have a cadence of 135 s \citep{sunetal2017} and we used 21 data sets from 17:22 - 18:07 UTC.

The chromospheric changes were derived from Interferometric Bidimensional Spectropolarimeter (IBIS) data from the DST \citep{cavallini2006,reardoncavallini2008}. The \ion{Ca}{2} 8542 \AA\ spectral line was observed with two different observing programs: a cadence of 56 s and a Ca line scan time of 18 s before 17:48 UT and subsequently a cadence of 37 s and a Ca line scan time of 15.5 s, and the fit was performed on data from 16:45-18:23 UT. To convert the measured chromospheric polarization into $B_{\rm LOS}$, only data with a sufficiently high polarization signal (max($V/I) \ge 2\%$) were considered and the weak-field approximation was applied. A more detailed description of the data reduction and fitting process can be found in \citet{kleint2017}.

 \begin{figure*}[tb] 
  \centering 
   \includegraphics[width=0.9\textwidth]{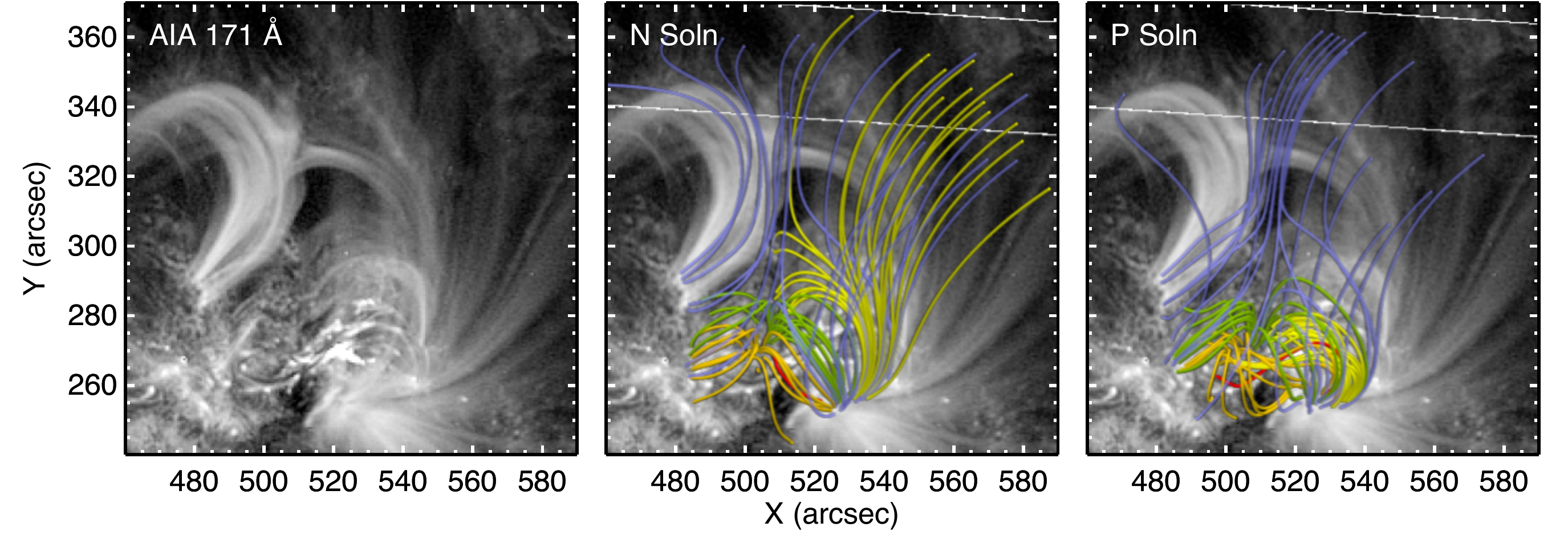}
   \caption{Two different magnetic field solutions obtained from the Grad-Rubin CFIT code and overlaid onto a contemporaneous 171 \AA{} SDO/AIA image. The field extrapolations are derived from the 29 March 2014, 17:36 UT, vector magnetogram data for AR 12017. CFIT reconstructs the coronal field using the values of the vertical magnetic field and force-free parameter $\alpha$ from the vector magnetogram as boundary conditions. The solutions presented here are obtained using values of $\alpha$ from points of negative magnetic polarity (N solutions, middle panel) and positive polarity (P solutions, right panel). The left panel shows just the SDO/AIA image, for comparison.}
        \label{npsols}
  \end{figure*}
  
NLFFF extrapolations require vector field data as input, which is only available in the photosphere. We used the HMI data series hmi.sharp\_cea\_720s and  hmi.B\_135s. 
The 720 s SHARP data contain only a part of the solar surface (usually one active region) and the magnetic field is retrieved by means of a Milne-Eddington inversion \citep[using the VFISV code,][]{borreroetal2011}. Because inversions contain a 180 degree ambiguity in the transverse magnetic field where $Q$ and $U$ cannot be uniquely assigned to a magnetic field direction, they are disambiguated in an additional step \citep[e.g.][]{metcalf1994}. The CEA in the data series stands for Lambert Cylindrical Equal-Area projection, which is a remapping of $\bm{B}$ from azimuth, inclination, and field strength into $B_{r}$ (the radial component), $B_{\phi}$ (the westward component of the field) and $B_{\theta}$ (the southward component of the field) \citep{xudongnote2013}. The 135 s vector data are a new product by the HMI team \citep{sunetal2017}. They are in HMI CCD coordinates and therefore we converted them to the CEA projection, which is suitable as input for the NLFFF code. We modified the routines \texttt{bvec2cea.pro} and \texttt{get\_bhzerr.pro} by Xudong Sun to replicate the output of the SHARP pipeline and verified our pipeline by comparing 135 s and 720 s data from the same timestamp, which agreed well in all observables. 

Extrapolations were performed with both the 135 s and 720 s data sets, to check whether the new 135 s data are comparable to the well-established 720 s data. In this paper we present only the NLFFF extrapolations from the higher cadence data, but we found that the results are the same with the 720 s data.

\section{NLFFF extrapolations}\label{nlfff}
We used the NLFFF code CFIT, which solves the NLFFF equations using the Grad-Rubin method \citep{wheatland07}. The NLFFF equations are replaced by linear equations, which are solved iteratively (by ``Grad-Rubin iteration''). The linear equations represent updates at a given iteration to the magnetic field in the computational volume and the electric current density in the volume. If the iteration sequence converges, the result is a solution to the NLFFF model (Equations~\ref{eq:nlfff1} and~\ref{eq:nlfff2}). 

\begin{figure*} 
  \centering 
   \includegraphics[width=.95\textwidth]{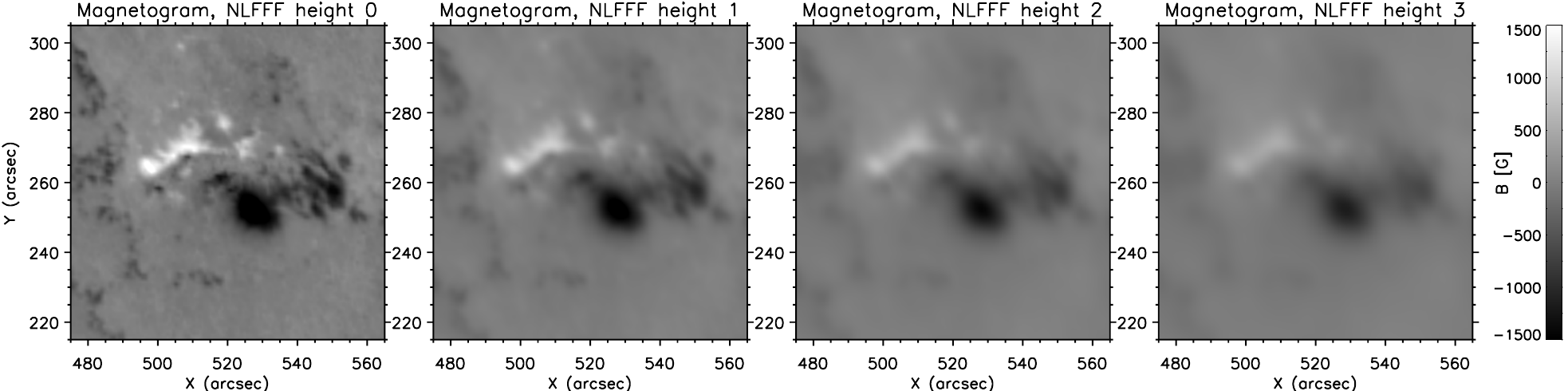}
   \caption{Comparison of line-of-sight magnetograms constructed for the NLFFF extrapolation at different heights, for the time 17:44:15 UT. Each height step is 725 km.}
        \label{heights}
  \end{figure*}

The CFIT code works in a cartesian geometry. The SHARP data are treated as field values on a cartesian grid with $B_x=B_{\phi}$, $B_y=-B_{\theta}$, and $B_z=B_r$, which corresponds to the lower boundary of the computational domain. The boundary conditions for the problem are the values of $B_z$ and the values of the force-free parameter $\alpha$ over one polarity of the field in the boundary (either the region where $B_z>0$ or the region where $B_z<0$)~\citep{wheatland07}. Values of $\alpha$ are obtained using $\alpha=\mu_0 J_z/B_z$, where $J_z$ is the local vertical component of the current density, which can be estimated from the values of $B_x$ and $B_y$ using finite differences. Hence the vector magnetogram data provide values of $\alpha$ over both polarities: they over-prescribe the problem. Two solutions can be constructed -- the P and the N solutions -- corresponding to the choice of values of $\alpha$ on either the positive or negative polarity. In practice, the two solutions may be quite different, because the vector magnetogram boundary data are inconsistent with the model \citep[e.g.][]{derosaetal09}. 

For the active region of interest here (AR 12017 on 29 March 2014) we find that the P and the N solutions are significantly different.  Figure~\ref{npsols} illustrates the problem. The middle panel shows the N solution obtained by CFIT for AR 12017 at 17:36 UT on 29 March 2014, and the right panel shows the P solution. Chosen field lines for the two solutions are shown, superposed on an aligned SDO Atmospheric Imaging Assembly (AIA; \citealt{lemenetal2012}) 171 \AA{} image, which indicates the structures in {the} corona. The left panel shows just the SDO/AIA image. The P solution features a highly twisted field structure (a flux rope) along the magnetic neutral line, which matches the location of a filament seen in the SDO/AIA image. However, this structure is absent in the N solution. {{This is because the photospheric positive polarity field has regions with high values of $\alpha$, which do not have counterparts in the negative polarity field. As a consequence of the force-free assumption, $\alpha$ is a constant along any field line in the NLFFF P solution, so the $\alpha$ values at the conjugate foot point (in the negative polarity) will not match the vector magnetogram values of $\alpha$. The origin of the discrepancy is that the photospheric field is forced, and hence inconsistent with the force-free model.}}

The energies of the two solutions are also different. We can define the free energy as the difference between total energy $E$ of the model field and the energy $E_0$ of the potential component of the model field. The free energy of the P solution of AR 12017, 29 March 2014 for 17:36 UT is $2.9\times 10^{31}$ erg (corresponding to $E/E_0=1.08$), and the free energy of the N solution at the same time is $9.0\times 10^{29}$ erg ($E/E_0=1.02$).

\citet{wheatlandregnier09} proposed the ``self-consistency procedure'' to address this problem. In this procedure, the P and N solutions are calculated, and then the different boundary values of $\alpha$ from the two solutions are averaged, subject to the uncertainties in the values. This provides a new set of boundary values of $\alpha$, from which new P and N solutions can be calculated (using also the common boundary values on $B_z$). The new solutions will again be different, but should be closer to agreement. This procedure is repeated (``cycled'')  until the P and N solutions agree. The final result is a single, self-consistent force-free solution \citep{wheatlandregnier09}.

 \begin{figure*}[tb] 
  \centering 
   \includegraphics[width=.8\textwidth]{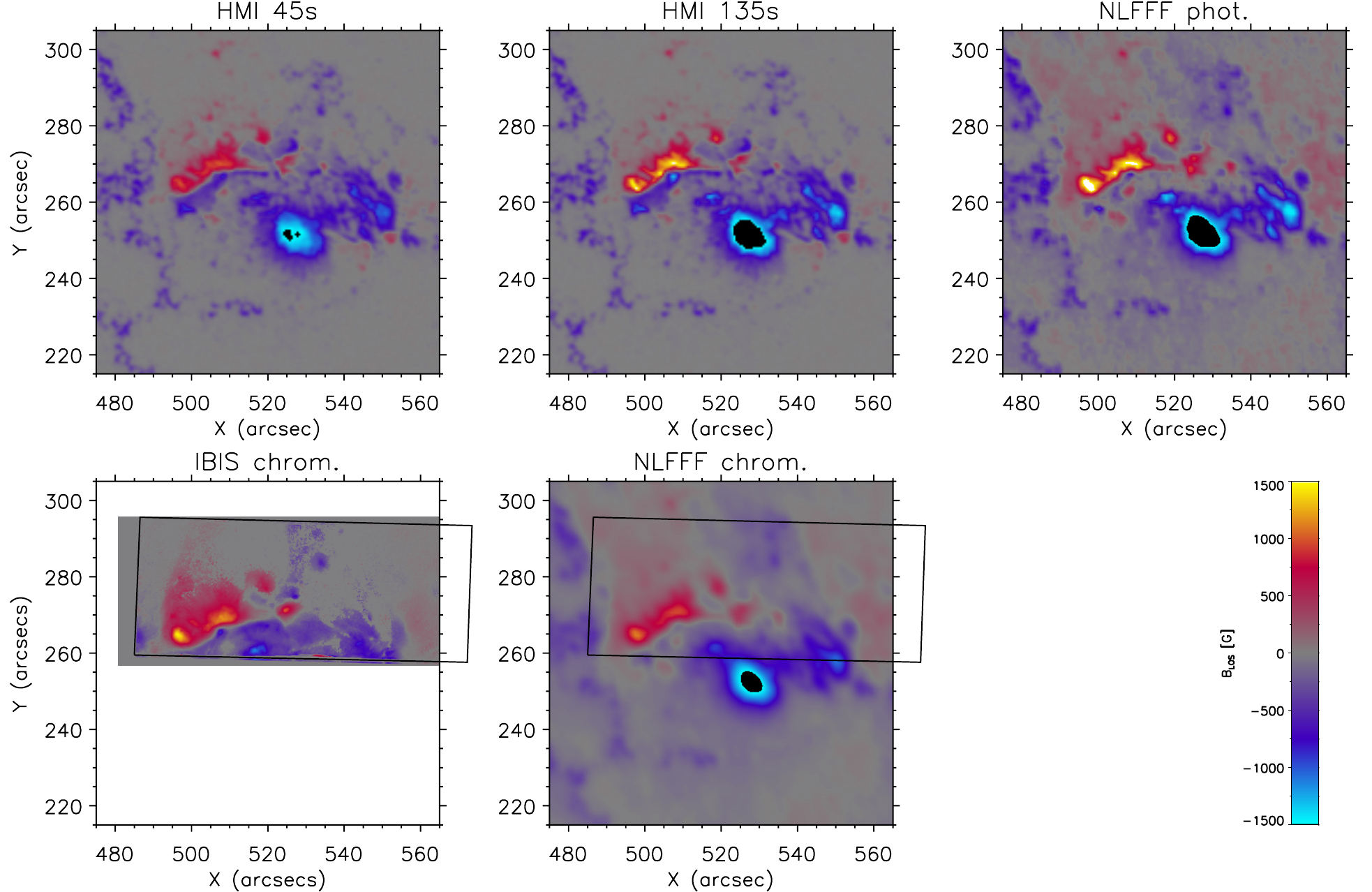}
   \caption{Comparison of line-of-sight magnetograms from the photosphere (top row) and the chromosphere (bottom row) for all data that were used for the analysis and the NLFFF extrapolations (see plot titles). The dataset closest to 17:45 UTC was selected for the comparison. The field of view of the IBIS instrument is drawn as a rectangle. It is known that 45 s magnetograms underestimate field strengths, but in all other cases the match is good.}
        \label{magnetograms}
  \end{figure*}

We apply CFIT, with self-consistency implemented, to the 2014 March 29, 17:15 -- 18:09 UT, 135s cadence vector magnetogram data from AR 12017. We rebin the data by a factor of 2 to speed up calculations, so that our volume is $346 \times 277 \times 115$ pixels in size, with each pixel being 1.005 arc seconds. In addition, on the boundary data we censor vertical currents with low signal-to-noise, \emph{i.e.} we set $\alpha=0$ at points with $\mathrm{SNR}(J_z)<1$, and we also set $\alpha = 0$ at points with $|B_z|<0.05\times \mathrm{max}(|B_z|)$. 
The number of self-consistency cycles required for CFIT to converge varies between each vector magnetogram boundary data set, but is usually $6$ -- $10$ cycles. 

We want to compare the field values in the NLFFF models with the photospheric (HMI) and chromospheric (IBIS) observations. To do this, we construct line-of-sight photospheric magnetograms from the NLFFF solution data cubes by taking the vector field values at the lower boundary and constructing the line-of-sight field component for the given viewing direction. For the chromospheric comparison, it is necessary to identify an appropriate height in the model.
Figure~\ref{heights} shows line-of-sight magnetograms constructed from the NLFFF extrapolation based on data at time 17:44:15 UT. The magnetograms are shown for the first few heights steps in the model, where height 0 is the photospheric boundary and where each step corresponds to 0.725 Mm. The Ca II 8542 \AA\ line forms in the low chromosphere from about 0.5 Mm (line wing) to $\approx$1.5 Mm (core). By comparing with the observed Ca II 8542 \AA\ magnetograms, we determined the first height step to match the observations most closely.

\section{Results}
\subsection{Comparison of Magnetograms}
First we compare the photospheric and chromospheric line-of-sight magnetograms from the available data and from the extrapolations. Figure~\ref{magnetograms} shows the photospheric magnetograms in the top row with the colors scaled between $-1500$ and $1500$ G. Values below and above these limits are indicated in black and white, respectively. The 45 s data, while qualitatively similar, underestimate the field strength, a known effect of the HMI pipeline \citep{hoeksemaetal2014}. The photospheric NLFFF data show more weak internetwork field than the original HMI data. To understand this discrepancy, we note that this active region was observed slightly off disk center ($\mu = 0.8$) and therefore the line-of-sight field $B_{\rm LOS}$ is a mixture of the values of $B_x$, $B_y$ and $B_z$ at the lower boundary in the NLFFF model. The boundary conditions for the NLFFF model are $B_{z}$ and $\alpha= J_z/B_{z}$ at the photosphere. The Grad-Rubin method preserves $B_{z}$ at the photosphere, but changes the values of $\alpha$, which means $B_{x}$ and B$_{y}$ at the photosphere change. This is the origin of the differences between the HMI line-of-sight magnetogram and the one constructed from the model. The bottom row shows the chromospheric IBIS data, whose FOV (indicated by the black rectangle) is limited by the telescope, and the first height step from the extrapolation. The weak field is invisible in IBIS because low polarization signals were excluded from the analysis. The  agreement between magnetograms from observations and extrapolations is generally good.

\begin{figure*} 
   \centering 
   \includegraphics[width=.94\textwidth,clip,bb=0 0 566 348]{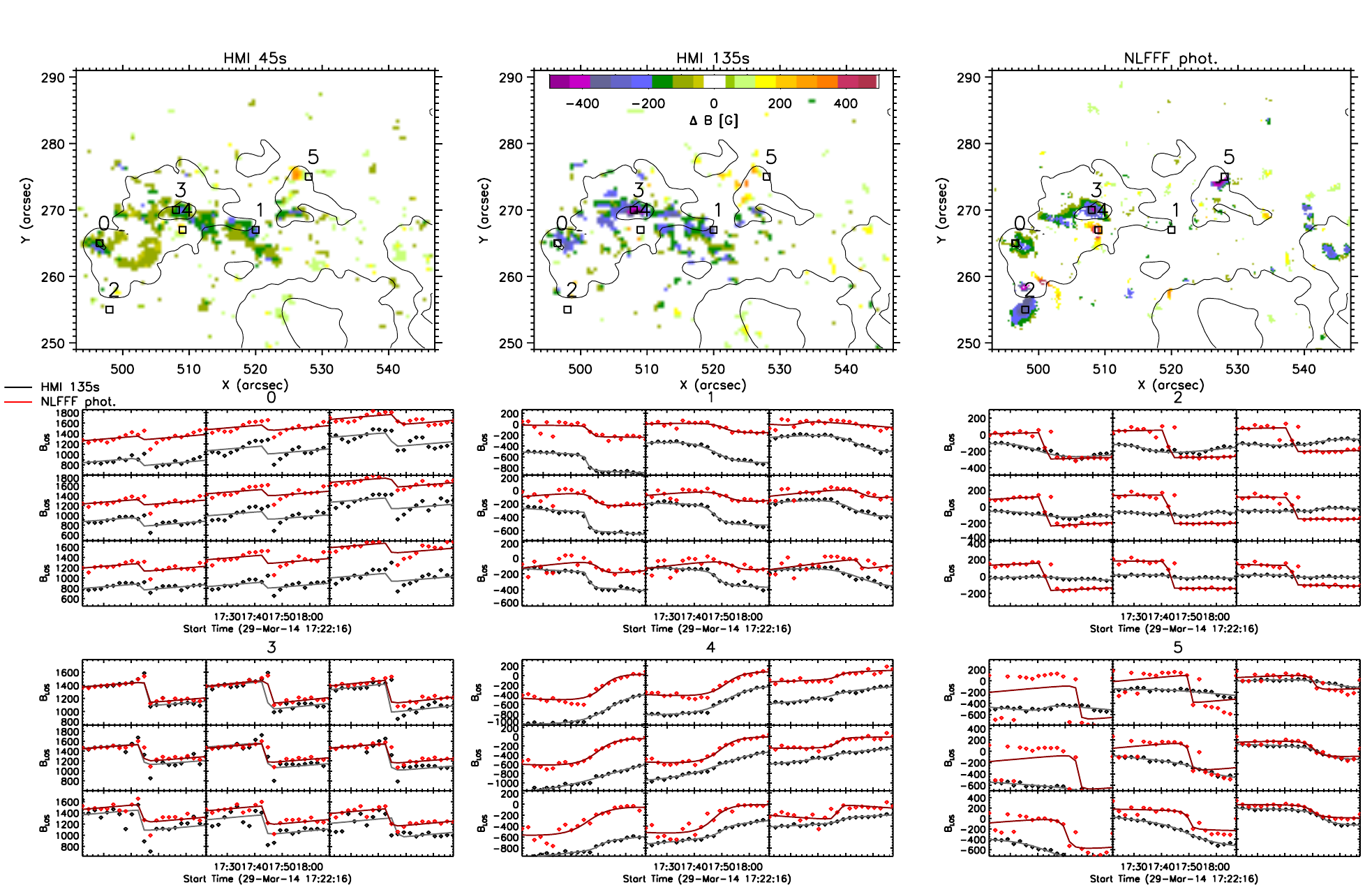}
   \caption{Comparison of stepwise magnetic field changes in the photosphere. The HMI 45 s data (top left) agree relatively well with the 135 s data (top middle), while the changes derived from the photospheric NLFFF sequence (top right) differ in some places. The lower panels show the evolution of $B_{\rm LOS}$ from the 135 s HMI data (black) and the photospheric NLFFF model (red) for a grid of 3$\times$3 pixels indicated by boxes in the upper images. The lines indicate the arctan fit. While many pixels agree, sometimes quantitatively and often qualitatively (e.g. positions 0, 1, 3, 4), the models include changes at locations that were constant during the flare (e.g. positions 2, 5).}
        \label{bchangephot}
  \end{figure*} 
\begin{figure*} %
   \centering 
   \includegraphics[width=.94\textwidth,clip,bb=0 0 566 348]{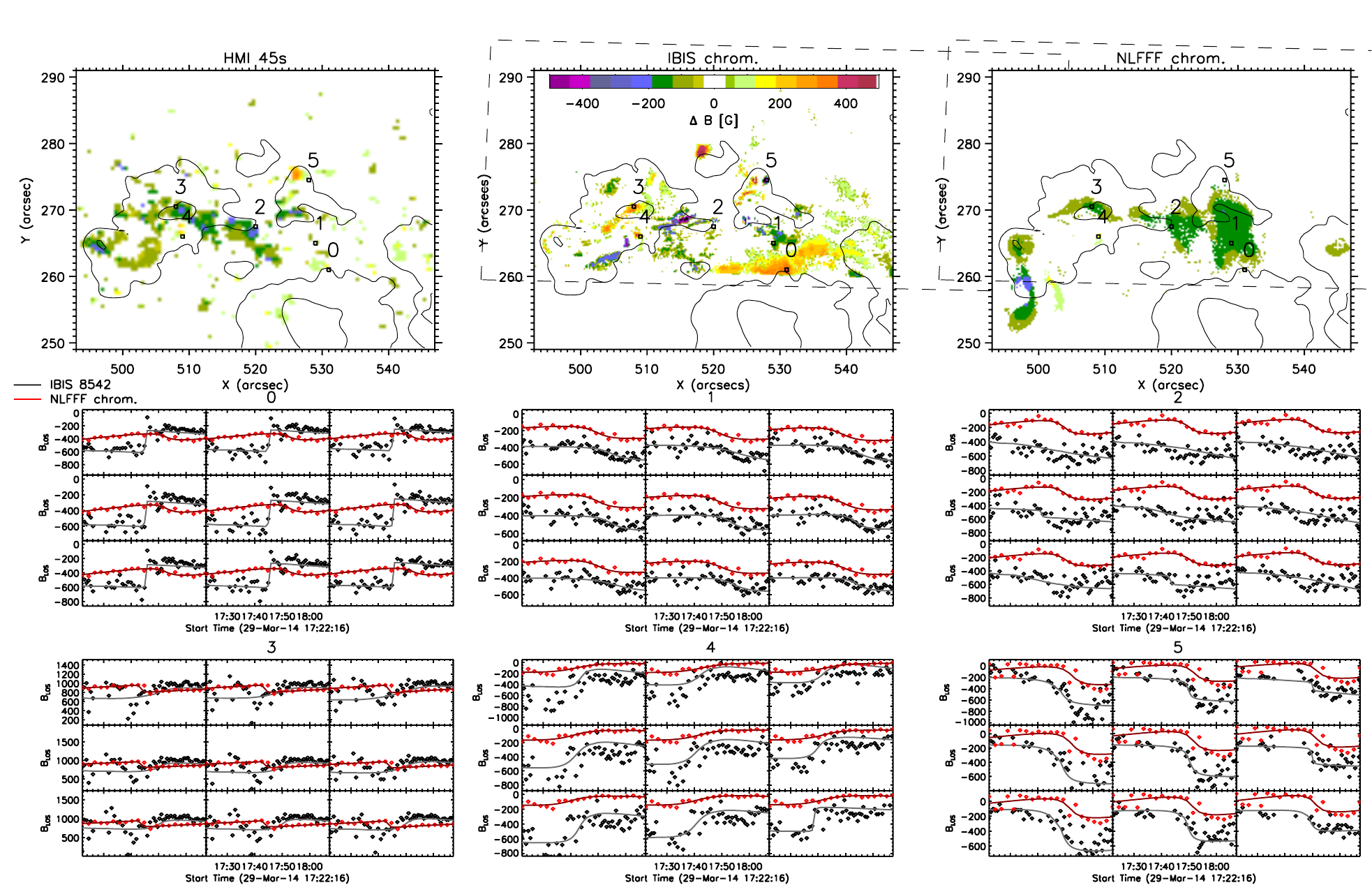}
   \caption{Comparison of stepwise magnetic field changes in the chromosphere (middle and right panels). The photospheric changes are shown for comparison in the left panel. In the chromosphere, there is no agreement between the observations and the NLFFF model.}
        \label{bchangechrom}
  \end{figure*}
  
\subsection{Comparison of Magnetic Field Changes}
Figure~\ref{bchangephot} shows the stepwise magnetic field changes in the photospheric observations and models. The top row shows the magnetic field changes (according to the color bar) at different spatial locations across the photosphere. To construct the panels in the top row, we verified the arctan fits for every pixel yielding steps above 80 G manually and excluded all data points below 80 G.
The HMI 45 s data are shown at the left of the top row (from \citealt{kleint2017}), the HMI 135 s data are in the middle, and the changes at the base of the NLFFF models are on the right. Five locations are also identified by boxes and numbers in each panel in the top row. The middle and bottom rows of the figure show the time variation of the field (in both the HMI 135 s and NLFFF photospheric data) for nine pixels at the location of each numbered box. The panels in the middle and bottom rows of Figure~\ref{bchangephot} also show the arctan fits used to determine the size of the change in field in each case.

The field changes seen in the HMI 45 s and 135 s data in the top row of Figure~\ref{bchangephot} agree well, at least qualitatively, although the size of the changes is larger in the 135 s data due to the previously mentioned underestimation of the field strength in the 45 s data. The photospheric NLFFF changes match the changes in the HMI data near the neutral line, which went through the eastern sunspot. For example box 0 shows a similar evolution of $B_{\rm LOS}$, only with a constant offset between observations and model. However, there are also locations where the HMI and NLFFF data disagree, for example boxes 2 and 5, where the NLFFF model shows a field change that was not observed on the Sun. {In order to generate a coronal field that is force-free and self-consistent, the NLFFF reconstruction process necessarily changes the horizontal magnetic field strengths at the photospheric boundary (while keeping the vertical magnetic field strength unchanged). The LOS magnetic field strengths of the NLFFF model are therefore different from those given by the magnetogram data, and the model field can show features and changes that are not observed in the data. The specific reasons for the discrepancies are difficult to identify.}
\begin{figure*}[!tbh] %
   \centering 
   \includegraphics[width=0.85\textwidth]{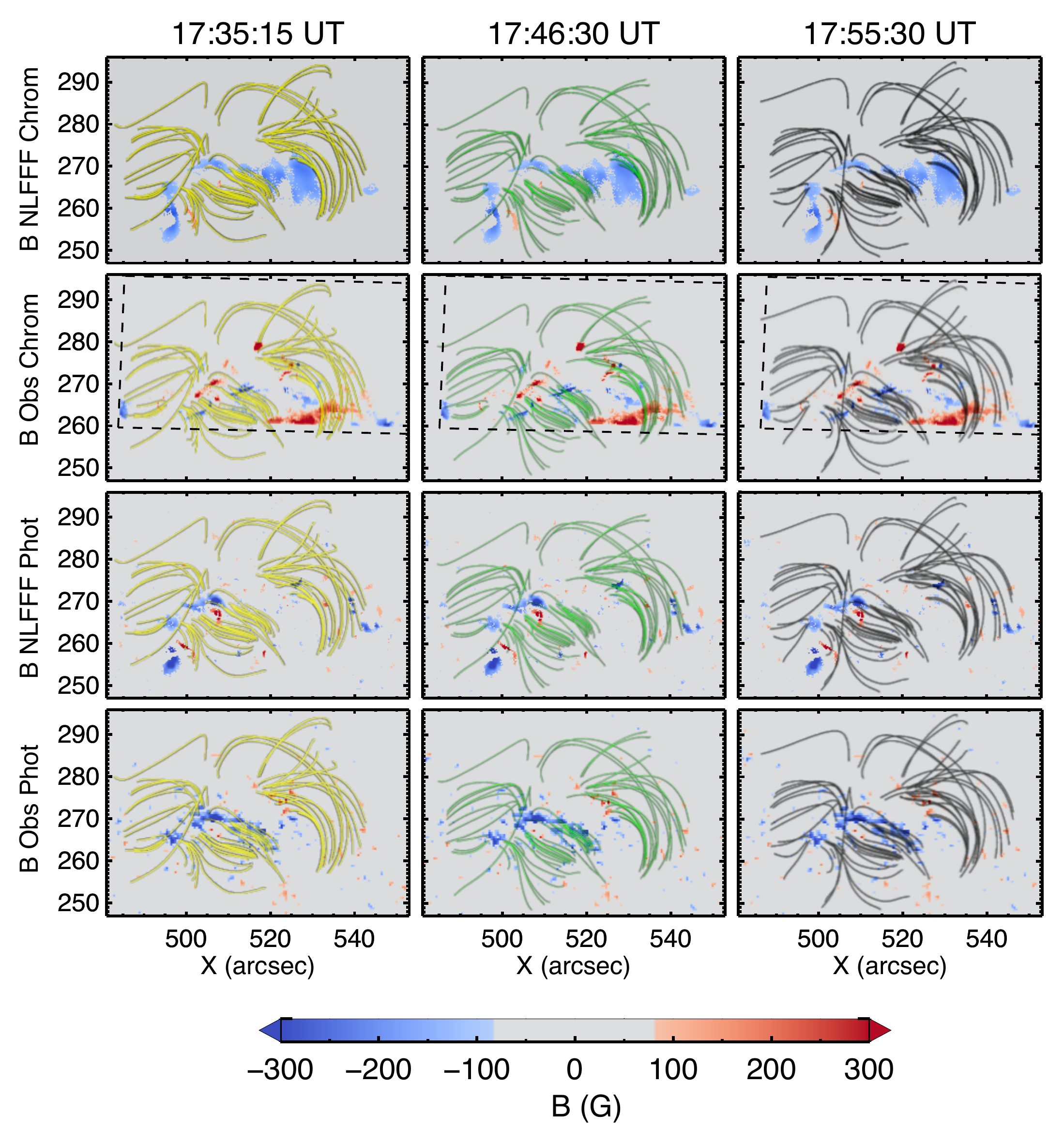}
   \caption{NLFFF model field lines overlaid onto the model and observational magnetic field changes. The field lines come from three specific times before, during, and after the flare, using fixed footpoint locations, while the $\Delta B_{\rm LOS}$ images reflect changes over the entire period for fixed locations on the Sun. The $\Delta B_{\rm LOS}$ images are thus the same for each column aside from small shifts to match the coordinates of the field-line extrapolations.}
        \label{overlay_all}
  \end{figure*}
  
The comparison of chromospheric changes is shown in Figure~\ref{bchangechrom}. The left panel in the top row shows the photospheric changes in the HMI 45 s data and is identical to the corresponding panel in the previous figure. The middle panel in the top row shows the changes derived from IBIS observations, and the right panel shows the changes at height index 1 (725 km height) in the NLFFF models. The poor match between model and observations is obvious. The model shows a broad region with a decrease in the line-of-sight field over the neutral line around [530, 265]\arcsec, which is only similar to the observations near box 1. The observations show a prominent region with a line-of-sight field increase centered around box 0 that is not reproduced by the NLFFF models.

To try to understand the observed (and NLFFF model) changes in relation to the model coronal magnetic field, we have overlaid field lines for the NLFFF solutions at three times on the diagrams showing the locations and magnitudes of permanent field changes. Figure~\ref{overlay_all} presents the results. The three columns correspond to three times: before (17:35:15 UT), during (17:46:30 UT), and after (17:55:30 UT) the flare. The first and second rows from the top show the changes in the NLFFF model and observations at the chromosphere. The third and fourth rows show the changes in the photosphere, in the same order. The field changes are shown in red when positive, and blue when negative.

Figure~\ref{overlay_all} illustrates that most of the changes, in the models and observations, occur around a set of low-lying loops in the models, which run along the neutral line. The observed chromospheric changes are predominantly associated with the footpoints of the loops, whereas the observed photospheric changes include a broad region of decreased line-of-sight field, which is centered around [510,270]\arcsec~and crosses the neutral line. This region underlies a set of very low loops in the models. The changes in the model at the photosphere do not reproduce this feature. The figure illustrates again the point made by \citet{kleint2017}, that the observed photospheric and chromospheric changes are very different. The observed changes have a complex relation with the model field lines. The observed changes at the chromosphere appear to include cases where the line-of-sight field increases at one footpoint and decreases at the other, and cases where the field increases at both footpoints.
  
In principle, a change in the line-of-sight field seen at the footpoints of coronal loops might be produced by a change in the orientation of the loops. If a field line at a positive footpoint tips towards the observer, the observed $B_{\rm LOS}$ increases, and if it tips away, the observed $B_{\rm LOS}$ decreases. We can estimate the implied change in angle as follows. If the magnetic field is at an angle $\theta_0$ to the line-of-sight and the angle changes by $\Delta \theta$ without changing the magnitude $B$ of the field, then the change in the line-of-sight field is
\begin{equation}
\begin{split}
\Delta B_{\rm LOS}&=B\cos(\theta_0+\Delta \theta)-B\cos\theta_0 \\
&=B\left(\cos\theta_0\cos\Delta\theta-\sin\theta_0\sin\Delta\theta-\cos\theta_0\right).
\end{split}
\end{equation}
Averaging over all possible angles $\theta_0$ gives
\begin{equation}
\langle\Delta B_{\rm LOS}\rangle = -\frac{2B}{\pi}\sin\Delta \theta.
\end{equation}
Taking $\langle \Delta B_{\rm LOS}\rangle\approx 200$ gauss and $B_{\rm LOS}\approx 1000$ gauss gives the change in angle $|\theta| = \sin^{-1} \frac{\pi}{2}\frac{\langle\Delta B_{\rm LOS}\rangle}{B}=18$ deg.

\begin{figure*}[tb] %
   \centering 
   \includegraphics[width=0.9\textwidth]{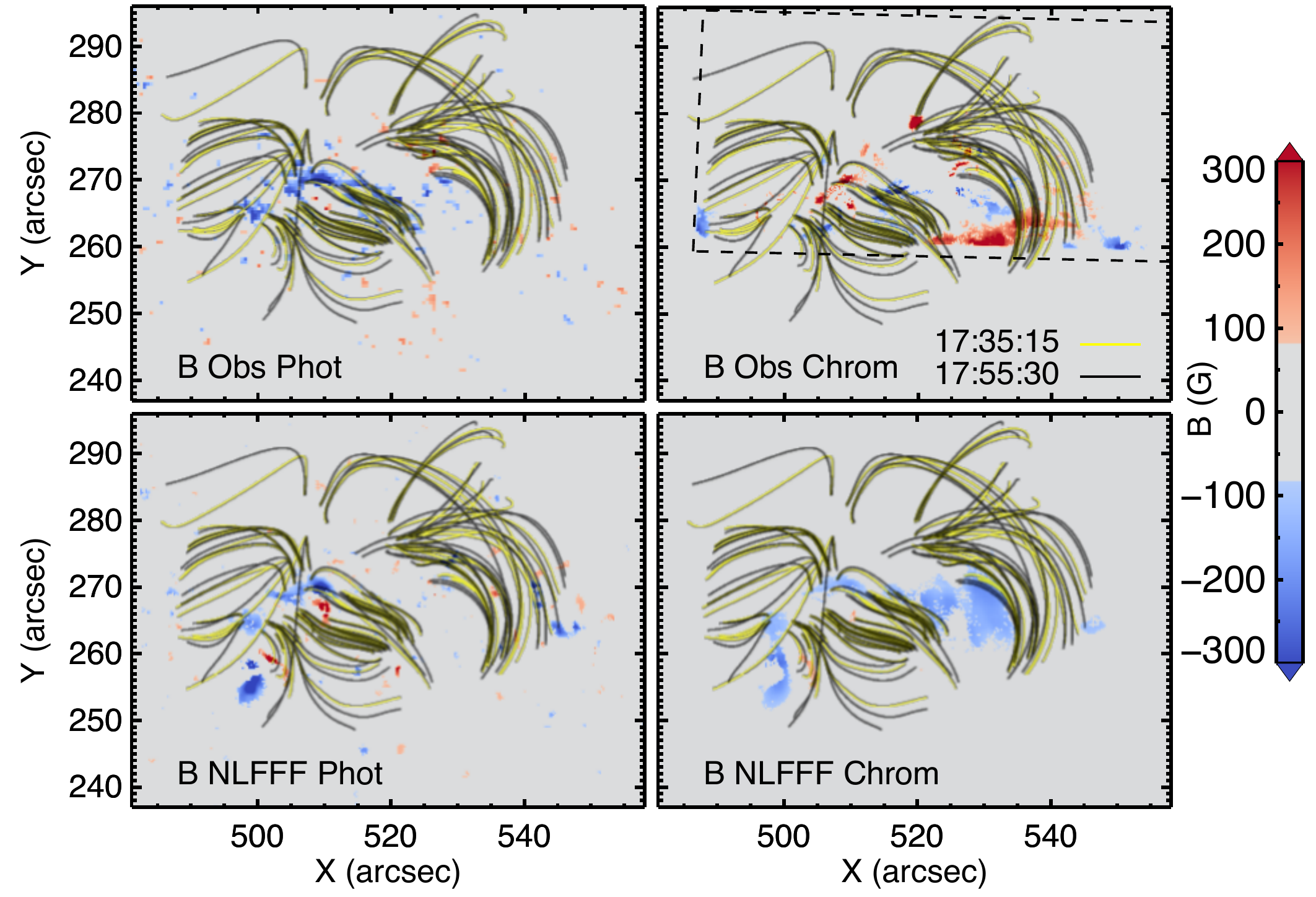}
   \caption{NLFFF model field lines before (yellow) and after (black) the flare overlaid onto the observed (top) and modeled (bottom) magnetic field changes in the photosphere (left) and chromosphere (right). Coordinates correspond to the later time (black, 17:55:30 UT), and the yellow field lines are shifted to match. The panels show that the field line orientation does not change drastically during the flare.}
        \label{overlay_comp}
  \end{figure*}

Figure~\ref{overlay_comp} illustrates the extent to which field lines in the NLFFF solutions change in direction. Each panel in the figure shows the NLFFF model field lines before (17:35:15 UT) and after (17:55:30 UT) the flare as yellow and black curves, respectively. The field lines are overlaid on images of $\Delta B_{\rm LOS}$, for the observations (top row) and for the models (bottom row). The left column shows the changes at the photosphere, and the right column shows the changes at the chromosphere. The figure shows clearly that the changes in orientation of the magnetic field lines between the before and after solutions are generally small. The changes in the line-of-sight field in the model involve changes in the magnitude of the field. To the extent that the changes in the model fields at the photosphere reproduce the changes in the HMI observations, this suggests that the observed field changes also involve a change in the magnitude of the field. However, as discussed, the modeled changes in the field at the chromosphere are quite different to those obtained from the IBIS observations.

\section{Discussion and Conclusions}

Our findings can be summarized as follows:
\begin{itemize}
\item Using an NLFFF model, we compared synthetic and observed line-of-sight magnetogram data to examine how well the model reproduces changes in the observed field strength at both photospheric and chromospheric heights over a $>$40 min period around the time of an X1-class flare. The models are constructed from HMI photospheric vector field data, but the model $B_{\rm LOS}$ is not identical to the photospheric observations because of how the NLFFF solutions are obtained. While there is generally good agreement between the model and observed magnetograms at both heights, there are significant differences in the locations and magnitudes of changes in $B_{\rm LOS}$.
\item The photospheric changes of the line-of-sight field $B_{\rm LOS}$ in the observations and NLFFF models match relatively well, especially near the neutral line.
\item The chromospheric changes in $B_{\rm LOS}$ in the NLFFF models and the observations do not agree. The observations show changes concentrated along the footpoints of loops that span the neutral line, while the NLFFF models indicate broader changes closer to the neutral line. The changes also do not match in sign.
\item The changes at the photosphere in the models (and to the extent that the models reproduce the data, the observations) are unlikely to be produced by a change in the orientation of the field alone. They involve changes in the magnitude of the field.
\end{itemize}

It is important to consider the influence of observational factors on the results. The observed chromospheric field changes appear to coincide with loop footpoints on either side of the neutral line, whereas the changes in the models are predominantly along the neutral line. This may be in part due to a reduced visibility of the \ion{Ca}{2} 8542 \AA\ line at the locations over the neutral line, where the field is nearly perpendicular to the line-of-sight. The influence of field configurations on the visibility might be tested by additional observations of flares with different locations on the disk, and also by forward modeling of the expected changes. Additionally, we are only considering a constant height in the model. During flares, the opacity of the atmosphere may change and we may be seeing different heights in the \ion{Ca}{2} 8542 \AA\ line. While we believe that this influence is not major because we fit a long time range of the observations and the intensity returns to pre-flare values during this time, we cannot fully exclude an influence. But it is known from simulations that the surface with an optical depth $\tau=1$ is corrugated even in non-flare cases, which means that our approximation of a constant height is not entirely accurate, but cannot be too far off because the observed and modeled magnetograms agree relatively well. 

Another observational factor is the method of determining the changes in the field associated with the flare. In principle, some of the inferred changes might be due to flux emergence or diffusion during the observing interval. However, in general these are slower processes and they are not expected to show clear jumps in 
$B_{\rm LOS}$ exactly at the flare time. This possibility was tested by performing the same arctan-fitting on data without any flares (the same region observed a few hours before and after this flare). The result was that no ``jumps'' larger than $150$ G were detected, and the number of small $B_{\rm LOS}$ changes was an order of magnitude lower than in the current sample. This provides confidence that most of the changes in this analysis are directly related to the flare.

The observed changes in the line-of-sight field at the photosphere and the chromosphere~\citet{kleint2017} are very different. {There is general agreement between the observed and model field changes at the photosphere, but discrepancy between the observed and model changes at the chromosphere. The likely explanation is that the model behavior at the chromosphere follows the photospheric data, and the NLFFF model excludes physics needed to reproduce the chromospheric changes. It is known that}
the magnetic connection between the photosphere and chromosphere is complex. The high-resolution chromospheric movie of the flare shows a very complex small-scale (subarcsecond) structure with changing loops, which do not seem to be reproduced by this global lower-resolution NLFFF model and this could be a major contributor to the observed discrepancies.
Tests of a nonlinear force-free reconstruction on boundary data from a radiative magneto-hydrodynamic simulation show that the NLFFF model performs poorly in representing the field structure in the chromosphere~\citep{2017ApJ...839...30F}. The low atmosphere is not force-free, and in particular the gas pressure and gravity force are dynamically important. The change in the magnetic pressure associated with the observed change in the line-of-sight field ($\Delta B_{\rm LOS}\approx 200$ gauss) assuming a line-of-sight field $B_{\rm LOS}\approx 1000$ gauss may be estimated in cgs units as $\Delta \left(B^2/8\pi\right)\approx \Delta (B_{\rm LOS}^2/8\pi)\approx B_{\rm LOS}\Delta B_{\rm LOS}/4\pi\approx 1.6\times 10^4$ dyne/cm$^2$. 
This is comparable to the gas pressure at a height of $\approx 250$ km in quiet-Sun models~\citep[e.g.][]{vernazzaetal1981}, and is much smaller than the photospheric gas pressure. On this basis, non-magnetic forces may play a role in the observed changes. However, these forces may be too slow to generate the large, stepwise changes observed on the flare time scale. The dynamic process most likely requires non-zero Lorentz forces.

In summary, for the purpose of reproducing the observed field changes at the photosphere and chromosphere, the force-free field model appears to be inadequate, because a NLFFF reconstruction based on photospheric boundary conditions does not include the physics of the chromosphere \citep{2017ApJ...839...30F}. A possible next step is a magneto-hydrostatic model \citep[e.g.][]{2016SoPh..291.3583G,2017SSRv..210..249W}, although this will require additional boundary conditions or simplifying assumptions. 

In conclusion the observed magnetic field changes in the X1 flare SOL20140329T17:48 remain a puzzle. Additional insight may also come from repeating the analysis for other well-observed flares.

\acknowledgments
This work was funded in part by an Australian Research Council Discovery Project (DP160102932). We thank Don Melrose for discussion and comments on the draft manuscript. IBIS is a project of INAF/OAA with additional contributions from the Universities of Florence and Rome and NSO.

\bibliographystyle{apj}
\bibliography{journals,ibisflare}

\end{document}